\def\Journal#1#2#3#4#5{{#1}, #2 {\bf #3,}  #4,  (#5)}

\def\Preprint#1#2{{#1}, #2}

\def\Nat{Nature}

\def\PRL{Phys. Rev. Lett.}
\def\APL{Appl. Phys. Lett.}
\def\JAP{J. Appl. Phys.}
\def\JLTP{J. Low Temp. Phys.}

\documentclass[prb,twocolumn,superscriptaddress,unsortedaddress,showpacs]{revtex4}
\usepackage{graphicx}
\usepackage{epsfig}

\begin{document}

\title{MgB$_2$ tunnel junctions and 19 K low-noise dc superconducting quantum interference devices}
\author{Y. Zhang}
\altaffiliation{Permanent address: Forschungszentrum Juelich, D-52425 Juelich, Germany}
\author{D. Kinion}
\author{J. Chen}
\affiliation{Department of Physics, University of California, and Materials Sciences Division,   Lawrence Berkeley National Laboratory, Berkeley, CA 94720}
\author{D. G. Hinks}
\author{G. W. Crabtree}
\affiliation{Materials Sciences Division, Argonne National Laboratory, Argonne, IL 90439}
\author{John Clarke}
\affiliation{Department of Physics, University of California, and Materials Sciences Division,   Lawrence Berkeley National Laboratory, Berkeley, CA 94720}

\date{\today}

\begin{abstract}
\vspace{0.3 cm}
Point contact junctions made from two pieces of MgB$_2$ can be adjusted to exhibit either superconductor-insulator-superconductor (SIS) or superconductor-normal metal-superconductor (SNS) current-voltage characteristics.  The SIS characteristics are in good agreement with the standard tunneling model for s-wave superconductors, and yield an energy gap of (2.02 $\pm$ 0.08) meV.  The SNS characteristics are in good agreement with the predictions of the resistively-shunted junction model.  DC Superconducting QUantum Interference Devices made from two SNS junctions yield magnetic flux and field noise as low as 4 $\mu \Phi _0$Hz$^{-1/2}$ and 35 fT Hz$^{-1/2}$ at 19 K; $\Phi _0$ is the flux quantum.
\end{abstract}


\maketitle

The discovery of superconductivity\cite{Nagamatsu2001} in MgB$_2$ at 39 K has generated considerable interest with regard to both fundamental issues and practical applications.  An important tool in both respects is the tunnel junction.  To measure the energy gap, a number of tunneling experiments with normal contacts have been performed.\cite{Karapetrov2001,Schmidt2001,Rubio2001,Kohen2001,Plecenik2001,Szabo2001,Sharoni2001}  These measurements reveal values of the low-temperature energy gap, $\Delta (0)$, that range from 2 to 7 meV; by contrast, the weak-coupling Bardeen-Cooper-Schrieffer (BCS) value $\Delta (0)$ = 1.76 $k_BT_c$ is 5.9 meV, where $T_c$ is the transition temperature.  Other experiments have been performed with both electrodes made of MgB$_2$.  Gonnelli {\em et al.}\cite{Gonnelli2001} used MgB$_2$ break junctions to observe a Josephson supercurrent, a nonhysteretic current-voltage (I-V) characteristic, and microwave-induced steps.  Brinkman {\em et al.},\cite{Brinkman2001} using nanobridges patterned in thin films of MgB$_2$, observed Josephson-like I-V characteristics and oscillations in the characteristics of dc Superconducting QUantum Interference Devices (SQUIDs) as a function of applied magnetic field.

In this Letter, we report experiments on all-MgB$_2$ point contact junctions that can produce either superconductor-insulator-superconductor (SIS) or superconductor-normal metal-superconductor (SNS) I-V characteristics.  The I-V characteristics of the SIS junctions are well fitted by the BCS tunneling model with a reduced energy gap.  The SNS junctions display I-V characteristics in good agreement with the resistively-shunted junction (RSJ) model,\cite{StewMcCumb1968} and are used to make dc SQUIDs with low noise at 19 K.

Compact samples of MgB$_2$ were formed from high purity amorphous B powder and Mg metal, and had a typical T$_c$ of 39 K.  Point contact junctions were made from flakes of MgB$_2$, typically 0.5 mm thick, with no further surface treatment.  The sharp point of one piece was pressed against a second with an adjustable screw arrangement.  The assembly was immersed in liquid $^4$He or raised above the bath to increase its temperature.  Three such junctions were made, and each was adjusted many times to obtain a wide variety of I-V characteristics.  

Figure 1 shows two representative four-terminal I-V characteristics for junctions with a resistance $R$ at high voltages ($\gg$ 2$\Delta /e$) of approximately 2.9 k$\Omega$.  The current is small for low voltages, and increases steeply at a voltage 2$\Delta /e$.  The absence of any discernible Josephson supercurrent is explained by the large values of the noise parameter, $\Gamma _N = 4eRk_BT/\Phi _0$ $\approx$ 0.33 and 0.67 in Figs. 1(a) and (b), respectively, which ensure that thermal fluctuations effectively quench the Josephson effect.\cite{Ambegaokar1969}  The curves through the data are fits to\cite{Dynes1978}

\begin{equation}
I = \frac{G_{NN}}{e} \int ^{\infty}_{- \infty} \rho (E) \rho (E + eV) [ f(E) - f(E + eV)] dE\text{,}
\end{equation}

\noindent
where G$_{NN}$  is the normal-state conductance fitted at voltages above 3$\Delta /e\text{, }\rho (E) = Re \left \{ (E-i\Gamma ) / [(E-i\Gamma )^2 - \Delta ^2]^{1/2}\right \}$ is the modified BCS density of states, $\Delta (T)$ is the fitted energy gap, $\Gamma (T)$ is a fitted gap-smearing parameter,\cite{Dynes1978} and $f(E)$ is the Fermi function.  Figure 1 also shows the differential conductances, $dI/dV$, obtained by differentiating the I-V curve, together with the theoretical prediction.  The peak around zero voltage arises from the gap smearing which produces an enhanced quasiparticle population at low energies.  The good agreement between theory and experiment confirms that tunneling occurs between two superconductors, rather than between a superconductor and a normal metal.

\begin{figure}
\centerline{\epsfig{file=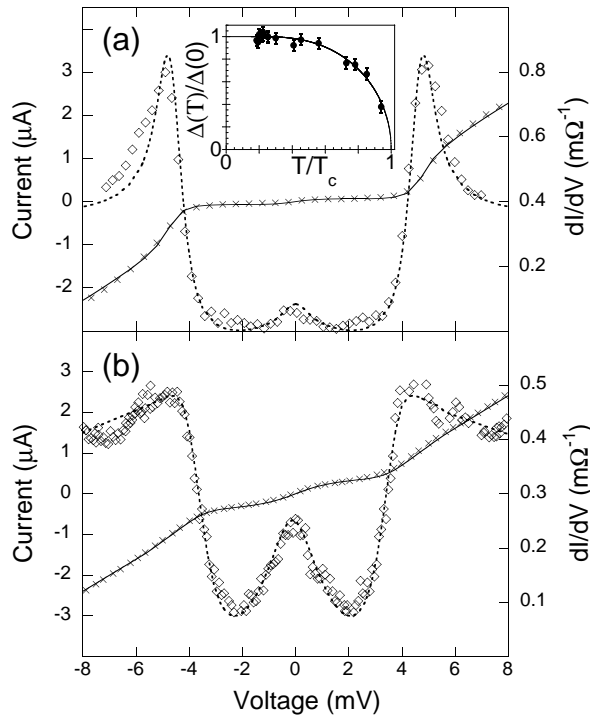,width=3.1 in,bbllx=59,bblly=302,bburx=432,bbury=763,clip=true}}
\caption{Current I (crosses) and conductance $dI/dV$ (diamonds) vs. voltage $V$ for MgB$_2$ tunnel junctions with fits to the theory shown as solid and dotted curves, respectively. (a) Temperature is 8.9 K, $\Delta$ = 2.06 meV, $\Gamma $ = 0.120 meV; (b) tempearture is 16.4 K, $\Delta$ = 1.88 meV, $\Gamma$ = 0.0469 meV.  Inset in (a) is $\Delta (T)$ vs. temperature $T$, fitted to BCS prediction with $\Delta (0)$ = (2.02 $\pm$ 0.08) meV and  $T_c$ = 29 K. }
\end{figure}

The inset in Fig. 1(a) shows $\Delta (T)$, extracted from a series of plots like those in Fig. 1, versus temperature.  Within the scatter in the data, the weak-coupling BCS prediction with a reduced gap (solid line) is a reasonably good fit.  However, the low temperature asymptote, $\Delta (0)$ = (2.02 $\pm$ 0.08) meV, is substantially below the value predicted by weak-coupling BCS theory and observed in some tunneling experiments to normal metal contacts,\cite{Karapetrov2001,Schmidt2001,Plecenik2001,Szabo2001,Sharoni2001} and the fit to the data indicates a $T_c$ of about 29 K.  These lowered values of $\Delta (0)$ and T$_c$ are possibly associated with a surface layer that has a reduced value of $T_c$ or is even normal, with a gap induced by the proximity effect.  However, the observed value of $\Delta (0)$ is still substantially below the BCS value predicted for $T_c$ = 29 K, about 4.4 meV.  

By increasing the pressure between the two MgB$_2$ surfaces, we obtain much lower resistances and nonhysteretic I-V characteristics with a Josephson supercurrent.  Figure 2 shows an example along with the fit to the prediction of the noise-free resistively shunted junction model, $V = (I^2 - I_0^2)^{1/2}R\text{; }I_0$ is the critical current.  The noise parameter, $\Gamma _N \approx$ 0.0021, is small and the noise rounding is minor.\cite{Dynes1978} The good fit indicates that there are no evident excess currents.  One inset in Fig. 2 shows microwave induced steps\cite{Gonnelli2001} at voltages $m\Phi _0f_m$ ($m$ = 0, $\pm$ 1, $\pm$ 2$\ldots$) induced by several values of microwave power at frequency $f_m$.  The second inset shows the existence of a critical current and steps at 35 K, well above the value of $T_c$ inferred from the temperature dependence of the energy gap in Fig. 1(a).  Possibly the SNS junctions have a much lower resistance because surface layers of the MgB$_2$ have been penetrated to reveal bulk MgB$_2$.

\begin{figure}
\centerline{\epsfig{file=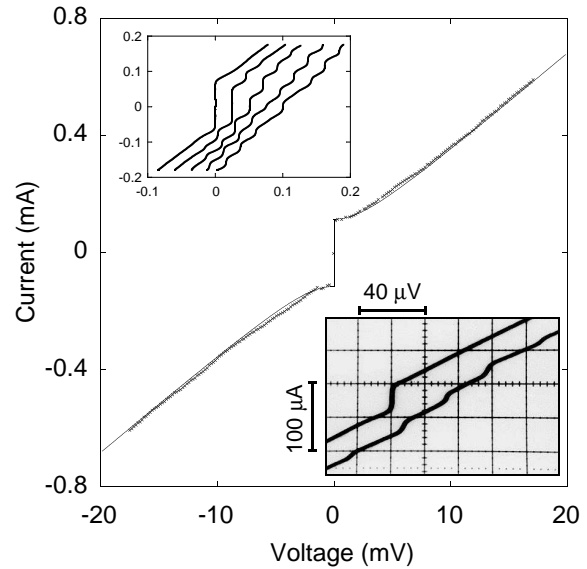,width=3.1 in,bbllx=36,bblly=335,bburx=467,bbury=763,clip=true}}
\caption{Current vs. voltage at 5 K for SNS junction with asymptotic resistance of 29.7 $\Omega$.  Curve is fit to noise-free resistively shunted junction model.  Upper inset shows current versus voltage at 20 K for increasing microwave power at frequency $f_m$ = 10 GHz; steps are at voltages $m\Phi _0f_m$, where $\Phi _0f_m$ = 20.7 $\mu$V.   Curves have been offset by 25 $\mu$V for clarity.  Lower right inset shows 10.3 GHz microwave-induced steps at 35 K.}
\end{figure}

We used the same technique to make dc SQUIDs.  The adjustable flake of MgB$_2$, typically 1.5 $\times$ 4.5 mm$^2$, was selected to have two nearby points that were pressed against the surface of the flat piece.  We attempted to make five SQUIDs in this way; three of them functioned well.  Figure 3 shows the I-V characteristic of one of them, with the magnetic flux $\Phi$ threading the loop adjusted to be $n\Phi _0 \text{  and }(n+1/2)\Phi _0$, where $n$ is an integer.  The inset in Fig. 3 shows the voltage versus applied magnetic flux. As a function of the bias current $I_B$, the peak-to-peak voltage of the oscillations peaks smoothly at a maximum of about 60 $\mu$V.  The oscillations are somewhat asymmetric, suggesting that the critical currents and resistances of the two junctions were unequal or the junctions were not placed symmetrically on the superconducting loop.\cite{Tesche1977}  At a bias current of 13.5 $\mu$A, the average maximum transfer coefficient, $V_{\Phi} = |\partial V/\partial \Phi|_{I_B}$, is approximately 280 $\mu \text{V}/\Phi _0$.

\begin{figure}
\centerline{\epsfig{file=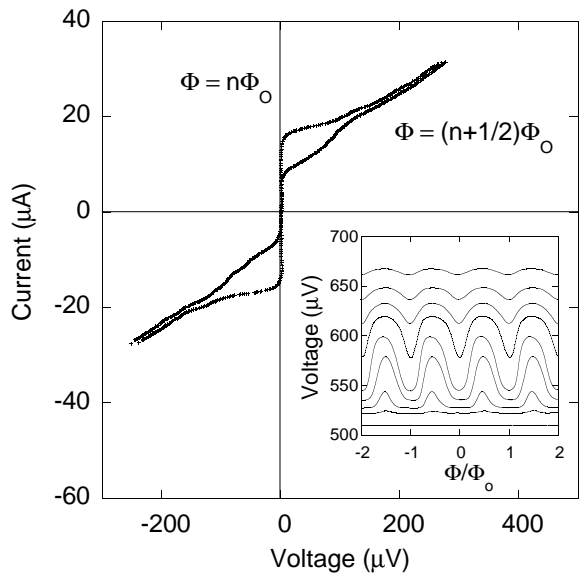,width=3.1 in,bbllx=88,bblly=245,bburx=511,bbury=671,clip=true}}
\caption{Characteristics of dc SQUID.  Main panel shows current versus voltage at 19 K for applied flux corresponding to integer and half-integer flux quanta.  Inset shows voltage versus applied magnetic flux for bias currents of 0, 6.6, 8.0, 11.0, 13.5, 16.0, 17.0, 18.0, and 19.2 $\mu$A.}
\end{figure}

We now make a rough comparison of the measured and predicted values of $V_{\Phi}$.  For the I-V characteristic in Fig. 3, the maximum critical current is 2$I_0 \approx \text{16} \mu$A, so that $\Gamma _N \approx$ 0.1 and the asymptotic resistance $R/2 \approx  11 \Omega$, where $R$ is the resistance (assumed to be equal) per junction.  The peak-to-peak swing in the critical current is $\Delta I_0 \approx 8 \mu$A, yielding $\Delta I_0/2I_0 \approx 1/2$.  Computer simulations\cite{Tesche1977} for ideal, identical Josephson junctions indicate that the corresponding value of $\beta _L \equiv 2LI_0 / \Phi _0 \approx $ 1, where $L$ is the inductance of the SQUID loop, implying that $L \approx $130 pH.  The predicted maximum value of the transfer coefficient is $V_{\Phi} \approx R/L \approx \text{ 350 }\mu V/\Phi _0$, in quite good agreement with the average measured value.

To measure the noise of our dc SQUIDs, we operated each in turn, surrounded by a high-permeability shield, in a flux-locked loop.  Figures 4(a) and 4(b) show the measured flux noise spectrum $S_{\Phi}^{1/2}(f)$ for two different SQUIDs at 19K. In Fig. 4(a), we observe a frequency independent (``white'') flux noise at frequencies down to about 500 Hz, with a value of about 4 $\mu \Phi _0 \text{Hz}^{-1/2}$.  As the frequency is lowered, the noise increases with a slope of roughly -1/4 down to about 3 Hz, and then increases steeply at lower frequencies.  The noise below 3 Hz is almost certainly due to ambient magnetic field fluctuations, and possibly due to mechanical instabilities in the point contact junctions.  The origin of the excess noise between 3 Hz and 500 Hz is less clear, and may or may not be intrinsic to the device.  The temperature of the SQUID was not well regulated, and fluctuations or drifts in temperature may well have contributed to the excess noise.  The right-hand ordinate shows the magnetic field noise, $S_B^{1/2}(f)  = S_{\Phi}^{1/2}/A_{eff}$; the effective area of the SQUID $A_{eff} \approx $ 0.16 mm$^2$ was found by measuring the magnetic field along the axis of the SQUID loop required to generate one flux quantum in the SQUID.  In the white noise region $S_B^{1/2}(f) \approx $ 50 fT Hz$^{-1/2}$.

\begin{figure}
\centerline{\epsfig{file=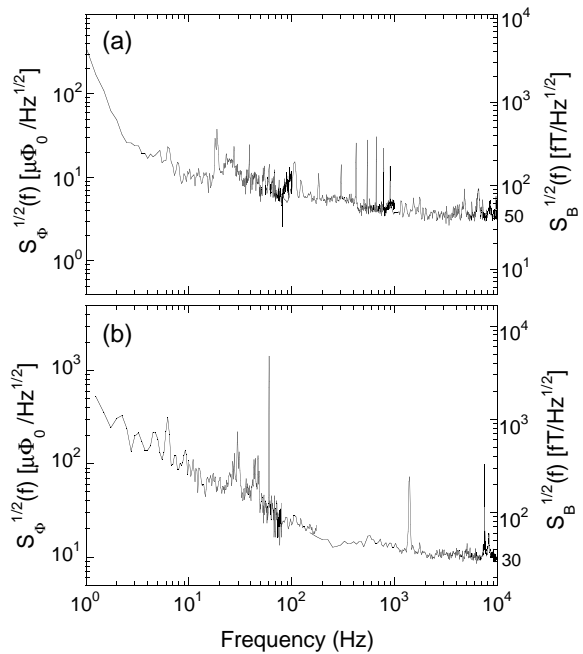,width=3.1 in,bbllx=94,bblly=180,bburx=570,bbury=720,clip=true}}
\caption{Noise spectra of two SQUIDs at 19 K.  Left-hand ordinate shows magnetic flux noise, right-hand shows magnetic field noise.}
\end{figure}

Figure 4(b) shows the noise for another SQUID.  The white noise, about 10 $\mu \Phi _0 \text{Hz}^{-1/2}$, extends down to about 1 kHz; $S_{\Phi}^{1/2}(f)$ scales approximately as $1/f$ at frequencies between about 100 Hz and 2 Hz.  The effective area $A_{eff}  \approx $ 0.60 mm$^2$, yielding a white magnetic field noise of 35 fT Hz$^{-1/2}$.  For both SQUIDs, operation of the flux-locked loop with bias current reversal\cite{Koch1983} - which reduces the low frequency noise due to fluctuations in the critical current - had no effect on the noise spectrum.

To compare the white flux noise for the SQUID in Fig. 4(a) with theory,\cite{Tesche1977} we estimate $\beta _L \approx $ 3, $L \approx $ 50 pH, $R \approx $ 20 $\Omega $ and $\Gamma _N \approx $ 0.04 to predict $S_{\Phi}^{1/2}(f) \approx (16k_BT/R)^{1/2}L \approx \text{ 0.4 } \mu \Phi _0 \text{Hz}^{-1/2}$.  A similar estimate for the SQUID in Fig. 4(b) yields 1$\mu \Phi _0 \text{Hz}^{-1/2}$.  The fact that the measured white noise for both devices is an order of magnitude higher than predicted may be due to asymmetric devices, poor radio-frequency shielding, and non-optimal matching of the SQUID resistance to the preamplifier.

Our observation of SIS-tunneling characteristics and the demonstration that thin films of MgB$_2$ can be deposited\cite{Brinkman2001} suggest that it may be feasible to fabricate thin film tunnel junctions with grown or deposited barriers. Such junctions could possibly extend the upper frequency ($\leq\text{ 2}\Delta/h$) at which SIS mixers\cite{Wengler1992} may be used.  In the case of the dc SQUIDs, the low-frequency noise  is 2-3 orders of magnitude lower than that of YBa$_2$Cu$_3$O$_{7-x}$ SQUIDs early in their development.  This result suggests that low frequency noise due to thermal activation of trapped flux vortices is less of an issue in MgB$_2$ SQUIDs than in their high-T$_c$ counterparts. 

This work was supported by the Director, Office of Science, Office of Basic Energy Services, Material Sciences Division of the U.S. Department of Energy under Contract Nos. DE-AC03-76SF00098 and W-31-109-ENG-38.

\end{document}